\documentclass[conference, a4paper]{IEEEtran}
\IEEEsettopmargin{t}{1.92cm}

\IEEEoverridecommandlockouts
\usepackage{cite}
\usepackage{amsmath,amssymb,amsfonts}
\usepackage{algorithmic}
\usepackage{graphicx}
\usepackage{textcomp}
\usepackage{xcolor}
\makeatletter
\let\MYcaption\@makecaption
\makeatother
\usepackage[font=footnotesize,labelformat=simple]{subcaption}

\makeatletter
\let\@makecaption\MYcaption
\makeatother
\usepackage{tikz}
\usepackage{comment}
\usepackage{pgfplots}
\usetikzlibrary{shapes,arrows,positioning,calc, matrix,spy,patterns}
\usepackage[nolist]{acronym}
\usetikzlibrary{external}
\pgfplotsset{compat=1.18}

\usepackage{booktabs}
\usepackage{xfrac}
\usepackage{comment}

\newcommand*{\figures}{figures}

\definecolor{cb-1}{HTML}{4477AA}
\definecolor{cb-2}{HTML}{EE6677}
\definecolor{cb-3}{HTML}{228833}
\definecolor{cb-4}{HTML}{CCBB44}
\definecolor{cb-5}{HTML}{66CCEE}
\definecolor{cb-6}{HTML}{AA3377}
\definecolor{cb-7}{HTML}{BBBBBB}

\pgfplotscreateplotcyclelist{cb list}{
	cb-1,cb-2,cb-3,cb-4,cb-5,cb-6,cb-7
}

\definecolor{kit-green100}{rgb}{0,.59,.51}
\definecolor{kit-green70}{rgb}{.3,.71,.65}
\definecolor{kit-green50}{rgb}{.50,.79,.75}
\definecolor{kit-green30}{rgb}{.69,.87,.85}
\definecolor{kit-green15}{rgb}{.85,.93,.93}
\definecolor{KITgreen}{rgb}{0,.59,.51}

\definecolor{KITpalegreen}{RGB}{130,190,60}
\colorlet{kit-maigreen100}{KITpalegreen}
\colorlet{kit-maigreen70}{KITpalegreen!70}
\colorlet{kit-maigreen50}{KITpalegreen!50}
\colorlet{kit-maigreen30}{KITpalegreen!30}
\colorlet{kit-maigreen15}{KITpalegreen!15}

\definecolor{KITblue}{rgb}{.27,.39,.66}
\definecolor{kit-blue100}{rgb}{.27,.39,.67}
\definecolor{kit-blue70}{rgb}{.49,.57,.76}
\definecolor{kit-blue50}{rgb}{.64,.69,.83}
\definecolor{kit-blue30}{rgb}{.78,.82,.9}
\definecolor{kit-blue15}{rgb}{.89,.91,.95}

\definecolor{KITyellow}{rgb}{.98,.89,0}
\definecolor{kit-yellow100}{cmyk}{0,.05,1,0}
\definecolor{kit-yellow70}{cmyk}{0,.035,.7,0}
\definecolor{kit-yellow50}{cmyk}{0,.025,.5,0}
\definecolor{kit-yellow30}{cmyk}{0,.015,.3,0}
\definecolor{kit-yellow15}{cmyk}{0,.0075,.15,0}

\definecolor{KITorange}{rgb}{.87,.60,.10}
\definecolor{kit-orange100}{cmyk}{0,.45,1,0}
\definecolor{kit-orange70}{cmyk}{0,.315,.7,0}
\definecolor{kit-orange50}{cmyk}{0,.225,.5,0}
\definecolor{kit-orange30}{cmyk}{0,.135,.3,0}
\definecolor{kit-orange15}{cmyk}{0,.0675,.15,0}

\definecolor{KITred}{rgb}{.63,.13,.13}
\definecolor{kit-red100}{cmyk}{.25,1,1,0}
\definecolor{kit-red70}{cmyk}{.175,.7,.7,0}
\definecolor{kit-red50}{cmyk}{.125,.5,.5,0}
\definecolor{kit-red30}{cmyk}{.075,.3,.3,0}
\definecolor{kit-red15}{cmyk}{.0375,.15,.15,0}

\definecolor{KITpurple}{RGB}{160,0,120}
\colorlet{kit-purple100}{KITpurple}
\colorlet{kit-purple70}{KITpurple!70}
\colorlet{kit-purple50}{KITpurple!50}
\colorlet{kit-purple30}{KITpurple!30}
\colorlet{kit-purple15}{KITpurple!15}

\definecolor{KITcyanblue}{RGB}{80,170,230}
\colorlet{kit-cyanblue100}{KITcyanblue}
\colorlet{kit-cyanblue70}{KITcyanblue!70}
\colorlet{kit-cyanblue50}{KITcyanblue!50}
\colorlet{kit-cyanblue30}{KITcyanblue!30}
\colorlet{kit-cyanblue15}{KITcyanblue!15}

\definecolor{cb-1}{HTML}{4477AA}
\definecolor{cb-2}{HTML}{EE6677}
\definecolor{cb-3}{HTML}{228833}
\definecolor{cb-4}{HTML}{CCBB44}
\definecolor{cb-5}{HTML}{66CCEE}
\definecolor{cb-6}{HTML}{AA3377}
\definecolor{cb-7}{HTML}{BBBBBB}

\pgfplotscreateplotcyclelist{cb list}{
	cb-1,cb-2,cb-3,cb-4,cb-5,cb-6,cb-7
}

\let\j\relax
\newcommand{\j}{\mathrm{j}}

\DeclareMathAlphabet{\mathbfsf}{\encodingdefault}{\sfdefault}{bx}{n}

\newcommand{\mvrv}[1]{\boldsymbol{\MakeLowercase{\mathsf{#1}}}}

\newcommand*{\vect}[1]{\boldsymbol{#1}}
\newcommand*{\mat}[1]{\MakeUppercase{\boldsymbol{#1}}}

\usepackage{ellipsis}
\usetikzlibrary{calc, patterns,fit,shapes}
\usetikzlibrary{decorations.pathreplacing,decorations.markings,shapes.geometric}
\tikzset{naming/.style={align=center,font=\small}}
\tikzset{antenna/.style={insert path={-- coordinate (ant#1) ++(0,0.25) -- +(135:0.25) + (0,0) -- +(45:0.25)}}}
\tikzset{station/.style={naming,draw,shape=dart,shape border rotate=90, minimum width=10mm, minimum height=10mm,outer sep=0pt,inner sep=3pt}}
\tikzset{mobile/.style={naming,draw,shape=rectangle,minimum width=12mm,minimum height=6mm, outer sep=0pt,inner sep=3pt}}
\tikzset{radiation/.style={{decorate,decoration={expanding waves,angle=90,segment length=4pt}}}}
\usetikzlibrary{decorations.pathreplacing}

\tikzset{pics/MUE/.style args={#1}
{code={
\node[mobile,label={[inner ysep=+.3333em]\dots}] (box){};
\draw ([xshift=.25cm] box.south west) circle (4pt)
      ([xshift=-.25cm]box.south east) circle (4pt);

\fill ([xshift=.25cm] box.south west) circle (1pt)
      ([xshift=-.25cm]box.south east) circle (1pt);

\draw ([xshift=.25cm] box.north west) [antenna=1];
\draw ([xshift=-.25cm]box.north east) [antenna=2];
}}}

\tikzset{pics/UE/.style args={#1}
{code={
\node[mobile] (box) {#1};

\draw ([xshift=.25cm] box.south west) circle (4pt)
      ([xshift=-.25cm]box.south east) circle (4pt);

\fill ([xshift=.25cm] box.south west) circle (1pt)
      ([xshift=-.25cm]box.south east) circle (1pt);

\draw (box.north) [antenna=1];
}}}

\tikzset{pics/MBS/.style args={#1}{code={
\node[station] (-base) {#1};
\draw[line join=bevel] (-base.100) -- (-base.80) -- (-base.110) -- (-base.70) -- (-base.north west) -- (-base.north east);
\draw[line join=bevel] (-base.100) -- (-base.70) (-base.110) -- (-base.north east);
\node (-a1) at ([xshift=.3cm,yshift=.3cm] -base.north) {};
\node (-a2) at ([xshift=-.3cm,yshift=.3cm] -base.north) {};
\draw[line cap=rect] ([xshift=.3cm,yshift=.3pt] -base.north) [antenna=1];
\draw[line cap=rect] ([yshift=.3pt]ant1 |- -base.north) -- ([xshift=-.3cm,yshift=.3pt] -base.north) [antenna=2];
\node (-d) at ([yshift=1cm] -base.east) {\dots};

}}}

\tikzset{pics/BS/.style args={#1}{code={

\node[station] (base) {#1};

\draw[line join=bevel] (base.100) -- (base.80) -- (base.110) -- (base.70) -- (base.north west) -- (base.north east);
\draw[line join=bevel] (base.100) -- (base.70) (base.110) -- (base.north east);

\draw[line cap=rect] ([yshift=0pt]base.north) [antenna=1];

}}}

\tikzset{pics/car/.style args={#1}{code={
    \node (-base) at (4.2,2){};
    \draw[very thick,#1, rounded corners=0.18ex,fill=black!20!blue!20!white]  (2.7,1.8) -- ++(1,0.7) -- ++(1.6,0) -- ++(0.6,-0.7) -- (2.7,1.8);
        \draw[thick,#1, fill=#1,thick] (4.2,2.5) -- (5.3,2.5)  -- ++(0.2,-0.23) -- (4.4,2.27) -- (4.4,1.8) --(4.2,1.8) -- cycle;
      \draw[thick]  (4.2,1.8) -- (4.2,2.5);
      \fill[draw=#1,fill=#1,rounded corners=0.6ex,very thick] (1.5,.5) -- ++(0,1) -- ++(1.2,0.3) --  ++(3,0) -- ++(1,0) -- ++(0,-1.3) -- (1.5,.5) -- cycle;
      \draw[draw=black,fill=gray!50,thick] (2.75,.5) circle (.5);
      \draw[draw=black,fill=gray!50,thick] (5.5,.5) circle (.5);
      \draw[draw=black,fill=gray!80,semithick] (2.75,.5) circle (.4);
      \draw[draw=black,fill=gray!80,semithick] (5.5,.5) circle (.4);
      \filldraw[draw=#1, fill=KITorange, semithick] (2.1,1.3) -- ++(150:.5) arc (155:205:0.5) -- cycle;
  
}}}

\tikzset{
  pobl/.style={
    inner sep=0pt, outer sep=0pt, fill=#1,
  },
  pobl gron/.style n args={2}{
    pobl=#1, rounded corners=#2,
  },
  pics/person/.style n args={3}{
    code={
      \node (-corff) [pobl=#1, minimum width=.25*#2, minimum height=.375*#2, rotate=#3, pic actions] {};
      \node (-pen) [minimum width=.3*#2, circle, pobl=#1, outer sep=.01*#2, anchor=south, rotate=#3, pic actions] at (-corff.north) {};
      \node (-coes dde) [pobl gron={#1}{1pt}, anchor=north west, minimum width=.12125*#2, minimum height=.25*#2, rotate=#3, pic actions] at (-corff.south west) {};
      \node [pobl=#1, anchor=north, minimum width=.12125*#2, minimum height=.15*#2, rotate=#3, pic actions] at (-coes dde.north) {};
      \node (-coes chwith) [pobl gron={#1}{1pt}, anchor=north east, minimum width=.12125*#2, minimum height=.25*#2, rotate=#3, pic actions] at (-corff.south east) {};
      \node [pobl=#1, anchor=north, minimum width=.12125*#2, minimum height=.15*#2, rotate=#3, pic actions] at (-coes chwith.north) {};
      \node (-braich dde) [pobl gron={#1}{.75pt}, minimum width=.075*#2, minimum height=.325*#2, outer sep=.0064*#2, anchor=north west, rotate=#3, pic actions] at (-corff.north east)  {};
      \node [pobl=#1, minimum width=.05*#2, minimum height=.2*#2, outer sep=.0064*#2, anchor=north west, rotate=#3, pic actions] at (-corff.north east) {};
      \node (-braich chwith) [pobl gron={#1}{.75pt}, minimum width=.075*#2, minimum height=.325*#2, outer sep=.0064*#2, anchor=north east, rotate=#3, pic actions] at (-corff.north west) {};
      \node [pobl=#1, minimum width=.0375*#2, minimum height=.2*#2, outer sep=.0064*#2, anchor=north east, rotate=#3, pic actions] at (-corff.north west) {};
      \node (-fit person) [fit={(-pen.north) (-braich dde.east) (-coes chwith.south) (-braich chwith.west)}] {};
      \node (-pwy) [below=25pt of -fit person, every pin] {\tikzpictext};
      \draw [every pin edge] (-fit person) -- (-pwy);
    };
  };
}

\tikzset{pics/phone/.style args={#1}{code={
    \node (-base) at (1,1.75) {};
      \draw[draw=black,fill=black!80,rounded corners=0.2ex,very thick] (0,0) -- ++(0,3.5) -- ++(2,0) --  ++(0,-3.5) -- ++(-2,0) -- cycle;
      \fill[thick, fill=KITblue] (0.2,0.2) -- ++(0,3.1)  -- ++(1.6,0) -- ++(0,-3.1) -- ++(-1.6,0) -- cycle;
      \fill[fill=#1] (0.65,2) rectangle (0.75,2.1);
      \fill[fill=#1] (0.85,2) rectangle (0.95,2.3);
      \fill[fill=#1] (1.05,2) rectangle (1.15,2.5);
      \fill[fill=#1] (1.25,2) rectangle (1.35,2.7);
  
}}}

\def\BibTeX{{\rm B\kern-.05em{\sc i\kern-.025em b}\kern-.08em
    T\kern-.1667em\lower.7ex\hbox{E}\kern-.125emX}}
\begin{document}
\begin{acronym}[TROLOLO]
  \acro{ACM}{auto-correlation matrix}
  \acro{ADC}{analog to digital converter}
  \acro{AE}{autoencoder}
  \acro{ASK}{amplitude shift keying}
  \acro{AoA}{angle of arrival}
  \acro{AWGN}{additive white Gaussian noise}
  \acro{BER}{bit error rate}
  \acro{BCE}{binary cross entropy}
  \acro{BICM}{bit-interleaved coded modulation}
  \acro{BMI}{bit-wise mutual information}
  \acro{BPSK}{binary phase shift keying}
  \acro{BP}{backpropagation}
  \acro{BSC}{binary symmetric channel}
  \acro{CAZAC}{constant amplitude zero autocorrelation waveform}
  \acro{CDF}{cumulative distribution function}
  \acro{CE}{cross entropy}
  \acro{CNN}{concolutional neural network}
  \acro{CP}{cyclic prefix}
  \acro{CRB}{Cramér-Rao bound}
  \acro{CRC}{cyclic redundancy check}
  \acro{CSI}{channel state information}
  \acro{DFT}{discrete Fourier transform}
  \acro{DNN}{deep neural network}
  \acro{DoA}{degree of arrival}
  \acro{DOCSIS}{data over cable services}
  \acro{DPSK}{differential phase shift keying}
  \acro{DSL}{digital subscriber line}
  \acro{DSP}{digital signal processing}
  \acro{DTFT}{discrete-time Fourier transform}
  \acro{DVB}{digital video broadcasting}
  \acro{ELU}{exponential linear unit}
  \acro{ESPRIT}{Estimation of Signal Parameter via Rotational Invariance Techniques}
  \acro{FEC}{forward error correction}
  \acro{FFNN}{feed-forward neural network}
  \acro{FFT}{fast Fourier transform}
  \acro{FIR}{finite impulse response}
  \acro{GD}{gradient descent}
  \acro{GF}{Galois field}
  \acro{GMM}{Gaussian mixture model}
  \acro{GMI}{generalized mutual information}
  \acro{ICI}{inter-channel interference}
  \acro{IDE}{integrated development environment}
  \acro{IDFT}{inverse discrete Fourier transform}
  \acro{IFFT}{inverse fast Fourier transform}
  \acro{IIR}{infinite impulse response}
  \acro{ISI}{inter-symbol interference}
  \acro{JCAS}{joint communication and sensing}
  \acro{KKT}{Karush-Kuhn-Tucker}
  \acro{kldiv}{Kullback-Leibler divergence}
  \acro{LDPC}{low-density parity-check}
  \acro{LLR}{log-likelihood ratio}
  \acro{LTE}{long-term evolution}
  \acro{LTI}{linear time-invariant}
  \acro{LR}{logistic regression}
  \acro{MAC}{multiply-accumulate}
  \acro{MAP}{maximum a posteriori}
  \acro{MLP}{multilayer perceptron}
  \acro{MLD}{maximum likelihood demapper}
  \acro{ML}{machine learning}
  \acro{MSE}{mean squared error}
  \acro{MLSE}{maximum-likelihood sequence estimation}
  \acro{MMSE}{miminum mean squared error}
  \acro{NN}{neural network}
  \acro{OFDM}{orthogonal frequency-division multiplexing}
  \acro{OLA}{overlap-add}
  \acro{PAPR}{peak-to-average-power ratio}
  \acro{PDF}{probability density function}
  \acro{pmf}{probability mass function}
  \acro{PSD}{power spectral density}
  \acro{PSK}{phase shift keying}
  \acro{QAM}{quadrature amplitude modulation}
  \acro{QPSK}{quadrature phase shift keying}
  \acro{radar}{radio detection and ranging}
  \acro{RC}{raised cosine}
  \acro{RCS}{radar cross section}
  \acro{RMSE}{root mean squared error}
  \acro{RNN}{recurrent neural network}
  \acro{ROM}{read-only memory}
  \acro{RRC}{root raised cosine}
  \acro{RV}{random variable}
  \acro{SER}{symbol error rate}
  \acro{SNR}{signal-to-noise ratio}
  \acro{SINR}{signal-to-noise-and-interference ratio}
  \acro{SPA}{sum-product algorithm}
  \acro{UE}{user equipment}
  \acro{VCS}{version control system}
  \acro{WLAN}{wireless local area network}
  \acro{WSS}{wide-sense stationary}
\end{acronym}

\title{Loss Design for Single-carrier Joint Communication and Neural Network-based Sensing
\thanks{This work has received funding 
	from the German
	Federal Ministry of Education and Research (BMBF) within the projects
	Open6GHub (grant agreement 16KISK010) and KOMSENS-6G (grant agreement 16KISK123).}

}

\author{\IEEEauthorblockN{Charlotte Muth, Benedikt Geiger, Daniel Gil Gaviria and Laurent Schmalen\\}

\IEEEauthorblockA{Communications Engineering Lab (CEL), Karlsruhe Institute of Technology (KIT)\\ 
		Hertzstr. 16, 76187 Karlsruhe, Germany, 
		Email: \texttt{\{first.last\}@kit.edu}\vspace*{-1ex}}
}

\maketitle

\begin{abstract}
We evaluate the influence of multi-snapshot sensing and varying \ac{SNR} on the overall performance of \ac{NN}-based \ac{JCAS} systems. To enhance the training behavior, we decouple the loss functions from the respective \ac{SNR} values and the number of sensing snapshots, using  bounds of the sensing performance. Pre-processing is done through conventional sensing signal processing steps on the inputs to the sensing \ac{NN}. The proposed method outperforms classical algorithms, such as a Neyman-Pearson-based power detector for object detection and ESPRIT for \ac{AoA} estimation for \ac{QAM} at low \acp{SNR}.
\end{abstract}


\acresetall
\section{Introduction}
Communication as well as sensing are vital services for our hyper-connected society. Sustainable and efficient solutions are extremely relevant in modern applications. An increase in spectral and energy efficiency is achieved by combining radio communication and sensing into one joint system instead of operating two separate systems. Therefore, this work focuses on the co-design of both functionalities in a single \ac{JCAS} system.
The future 6G network is expected to natively support \ac{JCAS} by extending the object detection to objects without communication capabilities, and performing general sensing
of the surroundings \cite{Wild2021}. With this approach, we expect to increase spectral efficiency by making spectral resources accessible for sensing while maintaining their use for communication, as well as an increase in energy efficiency because of the dual use of a joint waveform.

There is growing interest in data-driven approaches based on \ac{ML} since they can overcome deficits, such as hardware impairments, faced by algorithms based on model-based techniques \cite{MateosRamos2021, MateosRamos2023}. Algorithms including \ac{ML} are expected to be prevalent in 6G, as its use has matured in communication and radar processing \cite{Wild2021}.
\ac{ML} approaches are have been studied separately for communication systems \cite{OShea2017,Cammerer2020}, and in the context of radar \cite{JaraboAmores2008, Fuchs2020}. 
In \cite{MateosRamos2021,Muth2023}, an \ac{AE} for \ac{JCAS} in a single-carrier system has been proposed, performing close to a maximum a posteriori ratio test detector benchmark for single snapshot sensing of one radar target. 
The work of \cite{MateosRamos2023} extends these methods to an \ac{OFDM} waveform, which is a well-known technique to combine communication and radar \cite{Sturm2011, Braun2010}. They demonstrate the potential of deep-learning-based sensing to mitigate hardware mismatches. However, their research is limited to single snapshot estimation. Performing sensing on multiple snapshots should yield additional processing gains when targets do not move too fast.

In this paper, we study the monostatic sensing capabilities of a single-carrier wireless communication system with multiple snapshots. To reduce complexity, we investigate a single carrier transmission instead of multicarrier waveforms such as \ac{OFDM}. We analyze in detail impacts on communication and sensing in different \ac{SNR} environments and multi-snapshot sensing. Taking into account the different implications of missed and false detection of objects, a detector with a constant false alarm rate is designed using \acp{NN}. We show that the \ac{DNN}-based sensing can outperform classical benchmark algorithms, namely a Neyman-Pearson-based power detector and ESPRIT for the \ac{AoA} estimation.

\emph{Notation:} $\mathbb{R}$ and $\mathbb{C}$ denote the set of real and complex numbers, respectively. Sets are denoted by calligraphic letters, e.g., $\mathcal{X}$, with the cardinality of a set being $|\mathcal{X}|$. We denote vectors and matrices with boldface lowercase and uppercase letters, e.g. vector $\vect{x}$ and matrix $\mat{X}$. The element in the $n$-th row and $k$-th column of the matrix $\mat{X}$ is denoted as $x_{nk}$. The transpose and conjugate transpose of a matrix $\mat{X}$ are given by $\mat{X}^\top$ and $\mat{X}^H$ respectively, while the Hadamard product and the outer product are indicated with the operators $\odot$ and $\otimes$. The diagonal matrix $\mat{D}$ with diagonal entries $\vect{d}$ is denoted as $\text{diag}(\vect{d})$ and the all-one vector of length $N$ is denoted as $\boldsymbol{1}_N$.
A circular-symmetric complex normal distribution with mean $\mu$ and variance $\sigma^2$ is denoted as $\mathcal{CN}(\mu,\sigma^2)$. Random variables are denoted as sans-serif letters, e.g., $\mathsf{x}$,  multivariate random variables with boldface sans-serif letters (e.g., $\mvrv{x}$).  Mutual information and cross-entropy are denoted by  $I(\mathsf{x}_1,\mathsf{x}_2)$ and $H(\mathsf{x}_1 ||\mathsf{x}_2)$, respectively.
\begin{figure*}[t]
	\centerline{\input{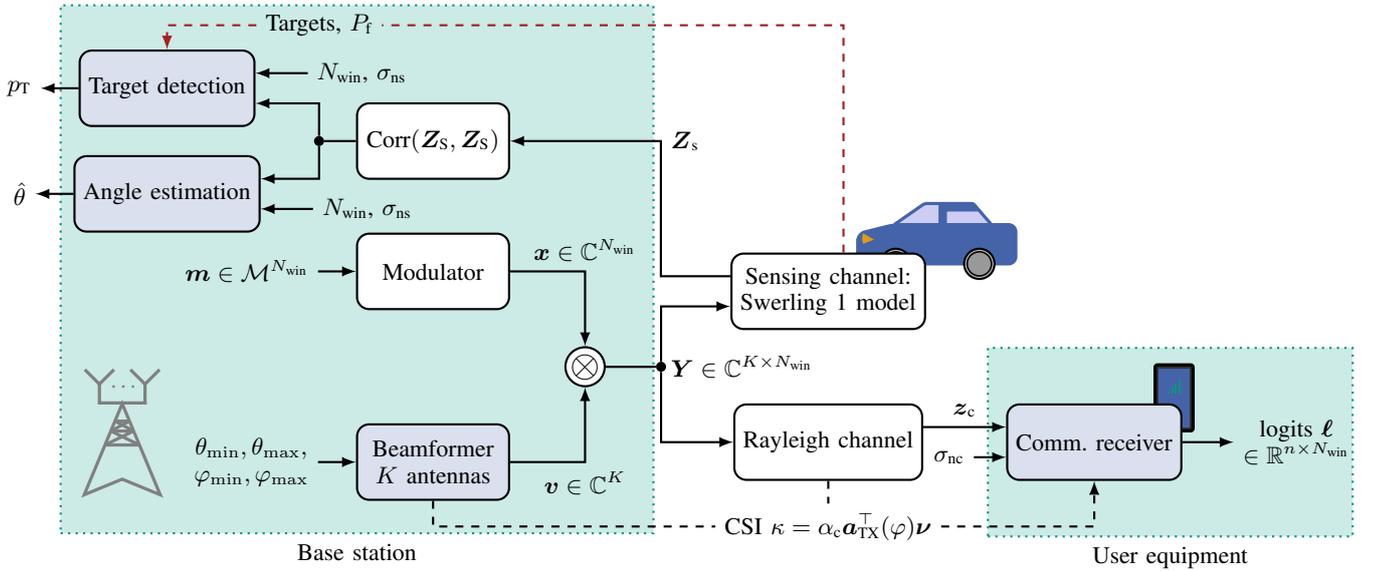}}
	\vspace{-0.5cm}
	\caption{\ac{JCAS} system, light blue blocks are trainable \acp{NN}, red dashed paths are only active while propagating the training data}
	\label{fig:flowgraphtrain_mono}
	\vspace*{-0.4cm} 
\end{figure*}
\section{System Model}\label{sec:sysmodel}

In this paper, we consider a monostatic \ac{JCAS} system, where the transmitter and the sensing receiver are co-located, i.e., part of the same base station, and are equipped with multiple antennas. Our objective is to detect a target and estimate the \ac{AoA} of the reflected signal in an area of interest. The transmit signal is simultaneously used to communicate with a \ac{UE} equipped with a single antenna in a \emph{different} area of interest. The communication receiver is located randomly at an azimuth angle $\varphi \in [\varphi_{\min},\varphi_{\max}]$ following an uniform probability distribution, and the \ac{AoA} of the radar target ${\theta}$ is drawn uniformly from $[\theta_{\min},\theta_{\max}]$. We consider multi-snapshot sensing with $N_{\text{win}}$ samples to provide more detailed information on multiple snapshot sensing.
The system block diagram is shown in Fig.~\ref{fig:flowgraphtrain_mono}. The blocks shaded in blue are realized as trainable \acp{NN}.

\subsection{Transmitter}
A modulator with $M$ different modulation symbols, while $M =2^n$, $n \in \mathbb{N}$, transforms the data symbols $m\nobreak \in \nobreak\mathcal{I}:=\nobreak \{1,2,\ldots,M\}$ into complex symbols $x \in \mathcal{M} \subset \mathbb{C}$. We generate a vector $\vect{x}\in \mathbb{C}^{N_{\text{win}}}$ of $N_{\text{win}}$ random symbols for block-wise processing. A fixed bit mapping maps a binary vector $\vect{b}=(b_1, \ldots, b_{n})^\top \in \{0,1\}^{n}$ to $m$. 

The second part of the transmitter generates digital precoding that causes beam directivity. For all $K$ transmit antennas, a unique complex factor $v_k = g_k\exp(\j\gamma_k)$ is generated for antenna $k \in \{1,2,\ldots,K\}$ with amplitude $g_k$ and phase shift $\gamma_k$ to steer the signal to our areas of interest. The beamformer inputs are the azimuth angle regions in which communication and sensing should take place, i.e. $\{\varphi_{\min},\varphi_{\max}, \theta_{\min},\theta_{\max}\}$. For block-wise processing, we consider a beamforming vector $\vect{v}\in \mathbb{C}^{K}$.
The modulator and beamformer employ power normalization to meet power constraints.

The transmit signal $\mat{Y}\in \mathbb{C}^{K \times N_{\text{win}}}$ is obtained by multiplying the complex modulation symbols $\vect{x}$ and the beamformer output $\vect{v}$, leading to
\begin{align}
	\mat{Y} = \vect{v} \vect{x}^{\top}.
\end{align}

\subsection{Channels}
A part of the transmit signal reaches the communication receiver while another part is reflected by the object of interest and reaches the sensing receiver co-located with the transmitter. 

For communications, the signal $\mat{y}$ experiences a single-tap Rayleigh channel before being received by the communication receiver with a single antenna as
\begin{align}
	\vect{z}_{\text{c}} =  \boldsymbol{1}^\top_K \left(\mat{a}_{\text{TX}}(\vect{\varphi}) \odot \mat{y}\right) \text{diag}(\vect{\alpha}_{\text{c}}) + \vect{n_{\text{c}}},
\end{align}
with complex normal distributed $\alpha_{\text{c},n} \sim \mathcal{CN}(0,\sigma_\text{c}^2)$ and $n_{\text{c},n}\nobreak \sim \nobreak \mathcal{CN}(0,\sigma_\text{n}^2)$. The signal propagation from $K$ antennas towards an azimuth angle $\varphi$ is modeled with the spatial angle matrix $\mat{a}_{\text{TX}}(\vect{\varphi})=(\vect{a}_{\text{TX}}(\varphi_{1})\, \ldots\, \vect{a}_{\text{TX}}(\varphi_{N_{\text{win}}})) \in \mathbb{C}^{K \times N_{\text{win}}}$ whose entries are given by
\begin{align}
	 \vect{a}_{\text{TX}}(\varphi) = \left(\exp\left(\j\pi  \sin\varphi \right),\ldots, \exp\left(\j\pi   K  \sin\varphi \right)\right)^\top,
\end{align}
assuming the antenna spacing to exactly match $\lambda/2$ of the transmission wavelength $\lambda$. The movement of the communication receiver leads to different $\varphi_n$ during an observation window.
We define $\text{SNR}_{\text{c}}= \frac{\sigma_\text{c}^2}{\sigma_\text{n}^2}$. The total \ac{SNR} needs to be corrected with the beamforming gain $\beta_{\text{c}}$ to $\text{SNR}=\nobreak \beta_{\text{c}} \cdot \text{SNR}_{\text{c}}$.

We express the sensing signal that is reflected from $T\in \nobreak \{0,1\}$ radar targets in the monostatic setup as
\begin{align}
	\mat{z}_{\text{s}} = T \vect{a}_{\text{RX}}({\theta}) \vect{a}_{\text{TX}}({\theta})^\top \mat{y} \text{diag}(\vect{\alpha}_{\text{s}}) + \mat{N_{\text{s}}},
\end{align}
with the radar target following a Swerling-1 model with ${\alpha}_{\text{s},n}\nobreak \sim \mathcal{CN}(0,\sigma_{\text{s}}^2)$ and $n_{\text{s},nk} \sim \mathcal{CN}(0,\sigma_{\text{n}}^2)$. The spatial angle vectors relate as $\vect{a}_{\text{RX}}({\theta})=\vect{a}_{\text{TX}}({\theta})$, with $\theta$ being the \ac{AoA} of the target throughout the observation window $N_{\text{win}}$. With a Swerling-1 model, we model scan-to-scan deviations of the \ac{RCS} that manifest as a change in $\alpha_{\text{s},n}$ during~$N_{\text{win}}$. The radial velocity of the target is assumed to be zero, so no Doppler shift occurs.

\subsection{Sensing Receiver}
At the sensing receiver, we want to detect a potential target and estimate its \ac{AoA} using a linear array of $K$ antennas in the base station. 
We consider multiple snapshot sensing with $N_{\text{win}}$ snapshots, enabled by forming the auto-correlation of all considered samples across the receive antennas
\begin{align}
    \text{Corr}(\mat{z}_{\text{s}},\mat{z}_{\text{s}}) := \frac{1}{N_{\text{win}}}  \mat{z}_{\text{s}} \mat{z}_{\text{s}}^H \quad \in \mathbb{C}^{K\times K}.
\end{align}
This signal is input to both the target detection and angle estimation blocks. The output of target detection is a probability value $p_{\text{T}} \in [0,1]$ that denotes the certainty that a target is present.
The angle estimation block outputs $\hat{{\theta}} \in [- \frac{\pi}{2}, \frac{\pi}{2}]$, which indicates the estimated azimuth \ac{AoA} of the target.

\subsection{Communication Receiver}
At the communication receiver, our goal is to recover information from the received signal.
We assume that the channel estimation has already been performed at the communication receiver and the precoding matrix $\vect{v}$ is known; therefore, \ac{CSI} $\vect{\kappa}=\nobreak \left(\mat{a}_{\text{TX}}^\top(\vect{\varphi})\vect{v}\right)\text{diag}(\vect{\alpha}_{\text{c}})$ is available at the communication receiver. It is important to note that this \ac{CSI} has no effect on the sensing functionality of the system. The receiver outputs are estimates of bitwise \acp{LLR} $\mat{L} \in \mathbb{R}^{n \times N_{\text{win}}}$ that can be used as input to a soft decision channel decoder. 

\subsection{Performance Indicators}
We formulate bounds on the communication throughput and the \ac{AoA} estimation accuracy as performance indicators for the system.
\subsubsection{Cramér-Rao Bound} 
For the estimation of the angle of a single target, the \ac{CRB} is given by
\begin{align}
\label{eq:CRB}
    C_{\text{CR}} (\theta) &= \frac{1}{\pi^2 \cos(\theta)^2}\frac{\sigma_{\text{ns}}^2}{2N_{\text{win}}}   \left[\frac{\sigma_{\text{ns}}^2 + K\beta \sigma_{\text{s}}^2}{K \beta^2 \sigma_{\text{s}}^3} \right] \cdot  \frac{6}{0.5K^3 - 0.5K},
\end{align}
where $N_{\text{win}}$ is the number of samples collected, $\theta$ the angle to be estimated, $\beta$ the beam-forming gain, and $K$ the number of antennas~\cite[Ch. 8.4]{Trees2002}. The given \ac{CRB} is a lower bound for the variance of an unbiased estimator.

\subsubsection{Bit-wise Mutual Information}
The goal of the communication receiver is to maximize the \ac{BMI}.
Maximizing the \ac{BMI} is equivalent to minimizing the \ac{BCE} between true bit labels $b_i$ and the estimated bit labels $\hat{b_i}$~\cite{Cammerer2020}. We can define it as:
\begin{align}
    \text{BMI}(\mvrv{B},\hat{\mvrv{B}}) &= \sum_{i=1}^{\log_2 M} I(\mathsf{b}_i; \hat{\mathsf{b}}_{i}).
\end{align} 
The \ac{BMI} is an achievable rate using binary coding and pragmatic coded modulation.

\section{Neural Network Training and Validation}

\subsection{Neural Network Configuration}\label{sec:NNconfig}
Our system is configured similarly to that of \cite{Muth2023} with adaptations described below and \ac{NN} layer dimensions given in Tab.~\ref{tab:NNsizes}. 
The \acp{NN} consist of fully connected layers with ELU activation function in the hidden layers. 
We implement modulation as a classical \ac{QAM}.
The inputs of the beamformer are the areas of interest for sensing with $\{\theta_{\min}, \theta_{\max}\}$ and for communication with~$\{\varphi_{\min}, \varphi_{\max}\}$. The output is subject to power normalization.
At the communication receiver, \ac{MMSE} equalization is performed, compensating for the random complex channel tap, to achieve better convergence along different \ac{SNR} values. The outputs of the communication receiver are interpreted as \acp{LLR} for each bit. For the \ac{BER} calculation, we use the hard decision of these \ac{LLR} values.

As stated in Sec.~\ref{sec:sysmodel}, we strive to improve the performance for multiple snapshot estimation by calculating the \ac{ACM} $\text{Corr}(\mat{Z}_{\text{s}},\mat{Z}_{\text{s}})$ of the different input matrices, similarly to the first processing step of the ESPRIT algorithm. We adapt the input layer of the sensing receiver to the new input dimension.
The number of input neurons of the sensing receiver is $K^2+2$.
Two input neurons have $N_{\text{win}}$ and $\sigma_{\text{ns}}$ as inputs and allow the investigation of sensing for varying channel parameters. Specifically, our systems are trained for generalized $N_{\text{win}}$ and $\sigma_{\text{ns}}$, allowing flexible investigation within the range of training parameters. This parameterization leads to roughly the same communication and sensing performance as systems trained individually for different $N_{\text{win}}$ and $\sigma_{\text{ns}}$, while allowing flexible operation without requiring a change of \ac{NN} weights and at the same time lowering the necessary computational complexity for training.

During training, the actual number of targets is fed to the sensing receiver in order to calculate the detection threshold that is needed to keep the false alarm rate $P_{\text{f}}$ constant. This threshold is added to the output of the detection \ac{NN} before applying the output sigmoid function. The output function of the angle estimation \ac{NN} is $\frac{\pi}{2}\tanh(\cdot)$, normalizing the output to $\pm \frac{\pi}{2}$.

\begin{table}[t]
\centering
\caption{Sizes of neural networks}
\begin{tabular}{@{}lccc@{}}
\toprule
component  & input layer & hidden layers &  output\\ 
\midrule
beamformer   & $4$      & $\{ K,K,2K \}$ & $2K$  \\
decoder   & $3$      & $\{10M,10M,10M,10M\}$          & $\log_2(M)$  \\
angle estimation   &   $2K^2+2$    & $\{8K,4K,4K,K\}$          & $1$  \\
detection   & $2K^2+2$       & $\{2K,2K,K\}$          & $1$  \\
\bottomrule
\end{tabular}
\label{tab:NNsizes}
\end{table}

\subsection{Loss Functions}
There are 3 main components of the loss function, resulting in a multi-objective optimization, which evaluates the performance of communication, detection, and angle estimation.
We introduce a weight $w_{\text{s}} \in [0,1]$, controlling the impact or perceived importance of the sensing tasks, resulting in the loss term
\begin{align}
    L = (1-w_{\text{s}})L_{\text{comm}} + w_{\text{s}}L_{\text{detect}} + w_{\text{s}}L_{\text{angle}} \label{eq:loss}.
\end{align}
\ac{JCAS} systems have been trained in~\cite{Muth2023} using the loss given by 
\begin{align}
    L = (1-w_{\text{s}})\underbrace{H(\mvrv{b}||\hat{\mvrv{b}})}_{L_{\text{comm}}} + w_{\text{s}}\underbrace{H(\mvrv{t}||\hat{\mvrv{t}})}_{L_{\text{detect}}} +  \frac{w_{\text{s}}}{N} \sum_{i=1}^{N} (\theta_i-\hat{\theta}_i)^2\label{eq:lossalt},
\end{align}
with $H(\mvrv{b}||\hat{\mvrv{b}})$ and $H(\mvrv{t}||\hat{\mvrv{t}})$ denoting the \ac{BCE} between transmitter and receiver for communication and target detection respectively.
While training multiple functionalities and multiple operating scenarios simultaneously, we observed a reduced performance when using a loss according to~\cite{Muth2023}. Especially the \ac{AoA} estimation showed unreliable convergence, since the achievable precision, which is bounded by~\eqref{eq:CRB}, depends significantly on the chosen $N_{\text{win}}$ and $\sigma_{\text{ns}}$. Therefore, the impact of trainable weights can be highly perturbed by these parameters. We introduce bound-informed adaptations to ensure good behavior over a range of \acp{SNR} and observation window lengths $N_{\text{win}}$.

In particular, we use the formulation of the \ac{CRB} in \eqref{eq:CRB} for an informed modification of the loss function used for the training of \acp{NN}.
Under the assumption of $\sigma_{\text{ns}}^2 \ll K\beta \sigma_{\text{s}}^2$, the factor ${\sigma_{\text{ns}}^2}/{N_{\text{win}}}$ describes the impact of the observation window and \ac{SNR} on the bound. We modify the loss term with the correction factor ${N_{\text{win}}}/{ \sigma_{\text{ns}}^2}$, achieving loss terms with similar magnitude for varying $N_{\text{win}}$ and $\sigma_{\text{ns}}$.
The proposed loss term is then given by:
\begin{align}
    L_{\text{angle}} =& \frac{1}{N} \sum_{i=1}^{N} \frac{N_{\text{win},i}}{ \sigma_{\text{ns},i}^2}(\theta_i-\hat{\theta_i})^2.\label{eq:anglelos}
\end{align}
The loss terms for detection $L_{\text{detect}}$ and communication $L_{\text{comm}}$ are equal to~\eqref{eq:lossalt}.

\subsection{Neural Network Training}\label{sec:nn-train}
The complete trained system is obtained in three phases: Pre-training, fine-tuning, and limiting. In pre-training, we first set $L_{\text{detect}}=0$ and in a secondary pre-training step $L_{\text{angle}}=0$ to initialize both sensing functionalities separately. 
We employ pre-training to establish the general behavior of the network. 
We use a total of $2.5\cdot\nobreak10^7$ communication symbols for both pre-training steps, divided into mini-batches of $10^4$ symbols. We use the Adam optimizer with a learning rate of $10^{-4}$. The length of the sensing window is randomly chosen between $1$ and $15$ for each sensing state, to optimize for different $N_{\text{win}}$ and give insight into the multi-snapshot behavior. Fine-tuning establishes the operating point of the \ac{JCAS} trade-off. Fine-tuning is performed on $5\cdot 10^7$ symbols by using the whole loss function in~\eqref{eq:loss} starting with the parameters established in the pre-training. We use the same hyperparameters as used for pre-training. Lastly, limiting ensures that the constant false alarm rate is kept. In the limiting phase, the system runs separately for $10^4$ symbols for each length of the sensing window $N_{\text{win}}$. The neural network components are not trained anymore in this phase, but the decision threshold for detection is refined as described in~\cite{Muth2023}.

For validation of the communication component, we choose the \ac{BER} as a metric.
The object detection task is evaluated on the basis of its detection rate and false alarm rate. For easier comparison, we design detectors with a constant false alarm rate.
The \ac{AoA} estimation is evaluated on the \ac{RMSE} of angle estimates of all targets. 

\section{Simulation Results and Discussion}\label{sec:sims}
In our simulations, the communication receiver is located at an \ac{AoA} of $\varphi \in [30^\circ,50^\circ]$. Radar targets are found in a range $\theta \in [-20^\circ,20^\circ]$.
Our monostatic transmitter and sensing receiver are both simulated as a linear array with $K=16$ antennas and we consider an observation window up to $N_{\max}=15$ and use a 16QAM as modulation format. For the radar receiver, our objective is to achieve a false alarm rate of $P_{\text{f}}=\nobreak10^{-2}$ while optimizing the detection rate and angle estimator.

\subsection{Benchmarks for Modulation, Detection and Angle Estimation}
\acused{ESPRIT}
As a benchmark detector, we employ a generalized power detector based on a Neyman-Pearson detector~\cite[Chap. 10]{Trees2002}, distinguishing between two normal distributions of mean $0$ with different variances. In the reference detector, the average power of all the input samples $z_{\text{s};il}$ considered for sensing is computed. The detector can be formulated as
\begin{align}
    \frac{2}{\sigma_{\text{ns}}^2}\sum_{l=1}^{N_{\text{win}}} \sum_{i=1}^{K} |z_{\text{s};il}|^2 \quad\mathop{\gtreqless}_{\hat{t}=0}^{\hat{t}=1}\quad \chi^2_{2KN_{\text{win}}}(1-P_{\text{f}}),
\end{align}
with the chi-squared distribution $\chi^2(\cdot)$ with parameter $2KN_{\text{win}}$ denoting the degrees of freedom of the distribution.
The correction factor is caused by the transformation of the problem from complex to real numbers, therefore artificially doubling the number of samples but reducing the noise by a factor of $1/\sqrt{2}$.
The benchmark detector has a constant false alarm rate along varying values of $\text{SNR}_{\text{s}}$ and $N_{\text{win}}$. 

We use the well-studied \ac{ESPRIT} algorithm as a benchmark for angle estimation as presented in~\cite{Trees2002}. \ac{ESPRIT} performs close to the \ac{CRB} for high \acp{SNR} or increasing observation windows $N_{\text{win}}$.%

\begin{figure}
\begin{subfigure}[c]{\columnwidth}
\hspace{0.5mm}
\includegraphics{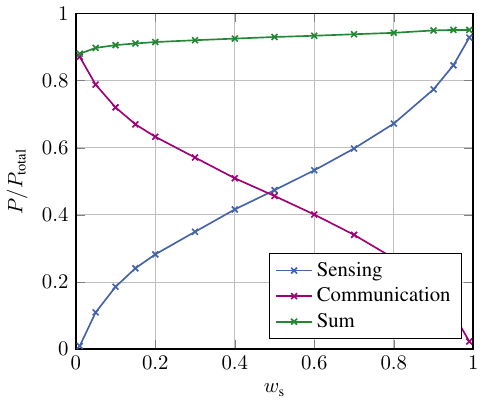}
\subcaption{Power distribution for communication receiver at $30\ldots50^\circ$ and sensing object at $-20\ldots 20^\circ$}
\label{fig:ws_comp}
\end{subfigure}\\
\begin{subfigure}[c]{\columnwidth}
\vspace{2mm}
\includegraphics{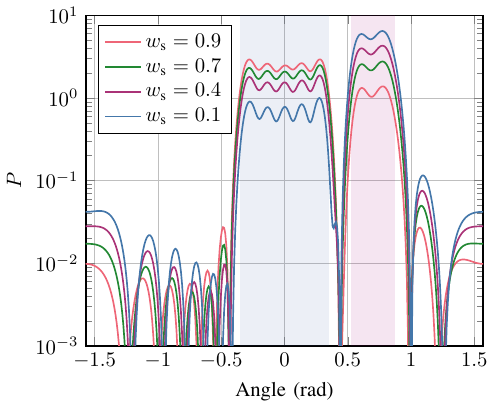}
\subcaption{Beam patterns outputs}
\label{fig:ws_comp_fine}
\end{subfigure}

\caption{Beamforming results for varying $\text{SNR}_{\text{c}}$ at different operating points $w_{\text{s}}$. The sensing area is filled light blue, the communication area light purple.}
\end{figure}

\begin{figure}
\includegraphics{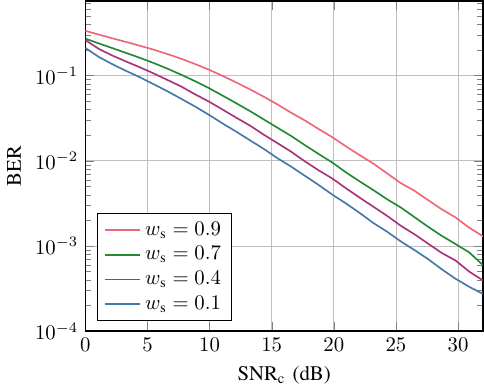}
 \vspace{-2mm}
\caption{Communication results for varying \ac{SNR}}
\label{fig:BER}
\vspace{-2mm}
\end{figure}

\subsection{Beamforming Results}
Figure~\ref{fig:ws_comp} shows the effect of the trade-off parameter $w_{\text{s}}$ on the power radiated to our areas of interest. We observe a linear relationship between $w_{\text{s}}$ and the power $P=\int_{\theta=\theta_{\min}}^{\theta_{\max}} \beta(\theta) \sigma_{\text{s}}^2 \mathrm{d} \theta$ in the region $[\theta_{\min},\theta_{\max}]$, which is closely related to our beamforming gain $\beta$. When $w_{\text{s}}=0.5$, the power is almost evenly distributed between sensing and communication. However, for $w_{\text{s}}<0.2$ and $w_{\text{s}}>0.8$, the trade-off becomes less linear, with the radiated power geared toward the less favored function decreasing more quickly. The total power radiated in both areas of interest increases slightly with increasing $w_{\text{s}}$, while around $10\%$ of the power is consistently radiated outside of our areas of interest. In Fig.~\ref{fig:ws_comp_fine}, the beam patterns for certain values of $w_{\text{s}}$ are displayed. The area for sensing is marked in light blue, while the area for communications is marked in light purple. We can see that most of the energy is radiated in our areas of interest and how the power is traded off between the two regions. We observe that more power is radiated outside our area of interest when only communication is performed, which is due to higher side lobes, particularly at an angle of $\pm 90^\circ$.

\begin{figure}
\includegraphics{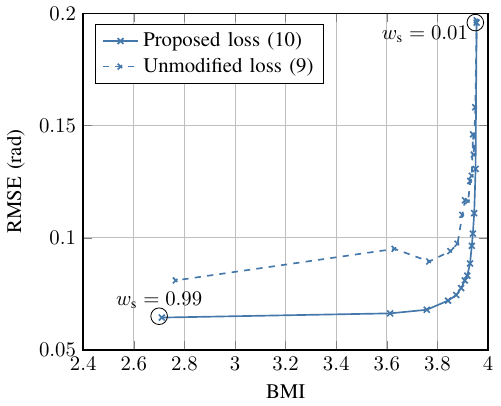}
\vspace{-2mm}
\caption{QAM operating at different $w_{\text{s}}$ after training with the modified loss function and without (dashed). Performance evaluated at $\text{SNR}_{\text{c}}=20.8\,\text{dB}$, $\text{SNR}_{\text{s}}=2.6\,\text{dB}$ and $N_{\text{win}}=1$}
\label{fig:BMI_vs_rmse}
\vspace{-2mm}
\end{figure}

\begin{figure*}
\begin{subfigure}[t]{0.32\textwidth}
\includegraphics{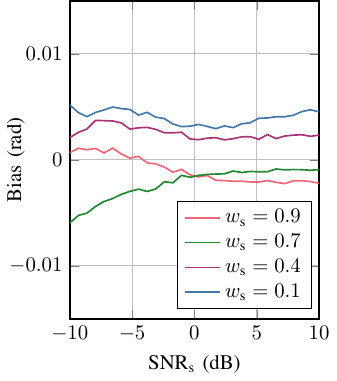}
	\vspace{-0.2cm}
	\subcaption{\ac{AoA} estimation bias for \ac{QAM} modulation}
	\label{fig:bias}
	\vspace{-0.2cm}
\end{subfigure}
\hspace{2mm}
\begin{subfigure}[t]{0.32\textwidth}
\includegraphics{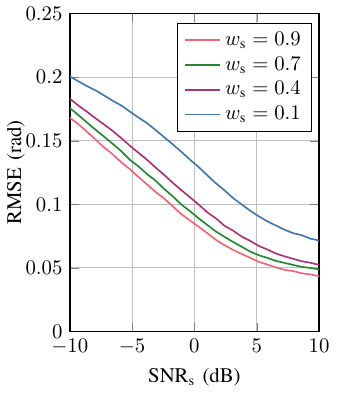}
        \vspace{-0.2cm}
	\subcaption{\ac{RMSE}}
	\label{fig:rmse_SNR}
\end{subfigure}
\hspace{2mm}
\begin{subfigure}[t]{0.32\textwidth}
\includegraphics{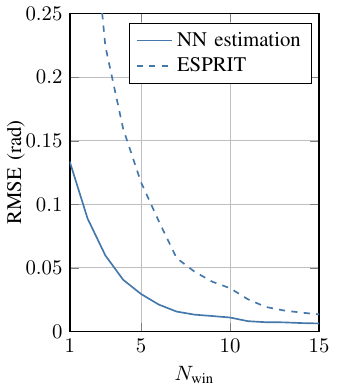}
	\vspace{-0.2cm}
	\subcaption{\ac{RMSE} with $\text{SNR}_{\text{s}} = -5\,$dB with $w_{\text{s}}=0.9$}
	\label{fig:rmse_cpr}
	\vspace{-0.2cm}
\end{subfigure}
\caption{\ac{AoA} estimation results evaluated on different \ac{SNR} for single snapshot sensing}
\vspace{-4mm}
\end{figure*} 

\begin{figure*}
\begin{subfigure}[t]{0.32\textwidth}
\includegraphics{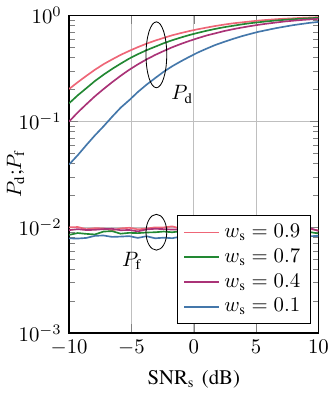}
	\vspace{-0.2cm}
	\subcaption{Single snapshot for different SNRs of the sensing channel with the NN modulation and QAM (dashed)}
	\label{fig:detection_SNR_NNQAM}
	\vspace{-0.2cm}
\end{subfigure}
\hspace{2mm}
\begin{subfigure}[t]{0.32\textwidth}
\includegraphics{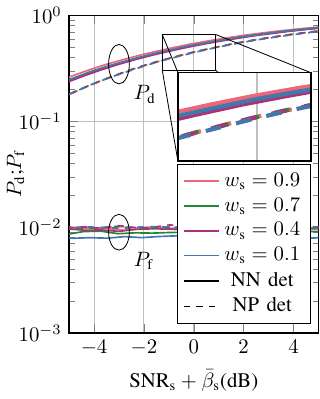}
	\vspace{-0.2cm}
	\subcaption{Single snapshot detection for QAM with Neyman-Pearson benchmark (dashed)}
	\label{fig:detection_SNRcorr_bm}
	\vspace{-0.2cm}
\end{subfigure}
\hspace{2mm}
\begin{subfigure}[t]{0.32\textwidth}
\includegraphics{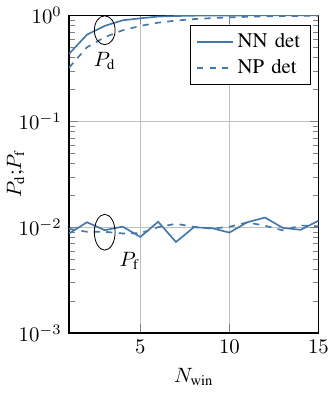}
	\vspace{-0.2cm}
	\caption{NN modulation and QAM with Neyman-Pearson benchmark, $\text{SNR}_{\text{s}}=-5\,$dB, $w_{\text{s}}=0.9$}
	\label{fig:pdpf_cpr}
	\vspace{-0.2cm}
\end{subfigure}
\caption{Detection probability and false alarm rate for varying \ac{SNR}}
\vspace{-4mm}
\end{figure*}

\subsection{Communication Results}
Previous works \cite{Cammerer2020, OShea2017} have demonstrated that \acp{NN} can be trained effectively as demappers. Figure~\ref{fig:BER} displays the \ac{BER} for a range of \ac{SNR} values and different trade-off factors $w_\mathrm{s}$.
As expected, we observe a higher \ac{BER} as function of the raw $\text{SNR}_{\text{c}}$ when increasing $w_{\text{s}}$. This degradation can be attributed to the beamforming gain towards the communication receiver, as the modulation format is identical for all $w_{\text{s}}$. 

\subsection{Angle Estimation Results}
We analyze the effect of the modified angle loss term on the results compared to the unmodified loss given in~\eqref{eq:lossalt} in Fig.~\ref{fig:BMI_vs_rmse}. We observe possible operating points of the \ac{JCAS} system that can be attained by selecting a different trade-off factor $w_{\text{s}}$. With the original loss function~\eqref{eq:lossalt}, as indicated by the dashed lines, the trained angle estimator demonstrates unsatisfactory performance. During training, we also noticed unreliable convergence of the system. We would anticipate a monotonic behavior of the curves, from high $w_{\text{s}}$ in the left bottom corner of the plot to low $w_{\text{s}}$ in the upper right corner. Without the modified loss function, there are performance fluctuations. The loss function is needed to optimize the neural network based on gradients with respect to the \ac{NN} weights. However, after training with the unmodified loss, the estimation error depends mainly on the noise power and $N_{\text{win}}$. Since $N_{\text{win}}$ and $\sigma_{\text{ns}}$ are randomly chosen from uniform distributions during training, the loss is influenced by random processes which distort the magnitude of the gradients. Nevertheless, this randomness during training is beneficial for generalization. By modifying the loss function, we keep the randomness in our training data while normalizing the expected magnitude of the gradients. We achieve the gradual performance trade-off as expected and achieve a lower \ac{RMSE} of the \ac{AoA} estimation than for the unmodified loss while a similar \ac{BMI} is achieved.

Results with a single target were already presented in \cite{MateosRamos2021, Muth2023}. 
In Fig.~\ref{fig:bias}, we analyze the estimation bias of the trained \ac{AoA} estimator. Since we use the formulation of the \ac{CRB} for an unbiased estimator, we should check for large biases that might cause unwanted effects in our system. The trained estimators lead to small biases in the order of $10^{-2}$. 
We do not observe a clear trend of the bias as a function of \ac{SNR} or $w_{\text{s}}$, nor do we observe a systematic bias.
In Fig.~\ref{fig:rmse_SNR}, we can observe how increasing $w_{\text{s}}$ decreases the angle estimation error for single snapshot sensing, as well as how the \ac{RMSE} decreases with increasing \ac{SNR}. The slope of the \ac{RMSE} flattens for higher \ac{SNR}.
in Fig.~\ref{fig:rmse_cpr}, we compare the \ac{RMSE} of \ac{AoA} estimation for different window lengths $N_{\text{win}}$. The trained angle estimators outperform ESPRIT at a raw $\text{SNR}_{\text{s}}=-5\,$dB. At low \ac{SNR}, the proposed method can consistently outperform the \ac{ESPRIT} baseline.

\subsection{Detection Results}
In Fig.~\ref{fig:detection_SNR_NNQAM}, we evaluate target detection with varying \ac{SNR}. The false alarm rate is kept approximately constant at $P_{\text{f}}\approx 10^{-2}$, as intended by design. Since we numerically calculate an appropriate decision threshold $T_{\text{off}}$ for each system, we observe small variations. The detection probability $P_{\text{d}}$ increases with $w_{\text{s}}$, indicating the impact of beamforming, especially for very low $w_{\text{s}}$, where the beamformer barely illuminates the sensing area. A detection rate of $0.5$ is obtained only at $\text{SNR}_{\text{s}}=2\,$dB for $w_{\text{s}}=0.1$. 

In Fig.~\ref{fig:detection_SNRcorr_bm}, we show the trained detectors together with the baseline detector but correct the \ac{SNR} with the average beamforming gain $\bar{\beta}_{\text{s}}$. We can observe that the benchmark detector leads to a lower detection rate, as it cannot take directional information into account. All detectors trained for \ac{QAM} perform almost identically.

In Fig.~\ref{fig:pdpf_cpr}, we evaluate the detection performance for different observation window lengths $N_{\text{win}}$. Increasing $N_{\text{win}}$ improves the detection rate. The false alarm rate remains approximately at $P_{\text{f}}$ for varying $N_{\text{win}}$ for all trained detectors, although it tends to vary more than for different \acp{SNR}. As we normalize the signal at the sensing receiver to the same noise floor, while numerically calculating separate thresholds for each $N_{\text{win}}$, this is expected. 

\section{Conclusion}\label{sec:concl}
\acused{ESPRIT}
In this paper, we have proposed a novel loss function for \ac{JCAS} systems based on supervised learning. By separating the loss function from the \ac{SNR} and observation window length for sensing, we have achieved a more reliable convergence of our system and improved its overall performance. We were able to adjust the trade-off between sensing and communication performance using the trade-off factor $w_{\text{s}}$. The trained object detector and \ac{AoA} estimator both outperform the baseline algorithms, namely a Neyman-Pearson-based power detector for object detection and ESPRIT for \ac{AoA} estimation.
Having reviewed the effects of multi-snapshot estimation encourages us not to use each communication sample to
perform sensing by itself in scenarios where objects are
\newpage%
\noindent slow enough to be captured by multiple samples in almost the same position. 
\vspace*{3mm}

\bibliography{IEEEabrv,literature_short.bib}
\bibliographystyle{IEEEtran}
\end{document}